\documentclass[conference]{IEEEtran}

\usepackage{graphicx}
\usepackage{cite}
	\usepackage[cmex10]{amsmath}
	\usepackage{mathabx}
\usepackage[ruled,vlined]{algorithm2e}
	\usepackage{array}
	\usepackage{mdwmath}
	\usepackage{mdwtab}
	\usepackage{eqparbox}
	\usepackage{url}
\usepackage{multirow}
\begin{document}

\title{\LARGE Target Privacy Preserving for Social Networks}


 \author{\authorblockN{Zhongyuan Jiang\authorrefmark{1}, Lichao Sun\authorrefmark{2}, Philip S. Yu\authorrefmark{2}, Hui Li\authorrefmark{1}, Jianfeng Ma\authorrefmark{1}, Yulong Shen\authorrefmark{1}
 }
 \authorblockA{\authorrefmark{1}Xidian University, Xi'an, Shaanxi 710071, China}
 \authorblockA{\authorrefmark{2}University of Illinois at Chicago, Chicago, IL 60607, USA \\
 zyjiang@xidian.edu.cn, \{lsun29, psyu\}@uic.edu}, hli@xidian.edu.cn, \{jfma, ylshen\}@mail.xidian.edu.cn}
\maketitle

\begin{abstract}
In this paper, we incorporate the realistic scenario of key protection into link privacy preserving and propose the target-link privacy preserving (TPP) model: target links referred to as \emph{targets} are the most important and sensitive objectives that would be intentionally attacked by adversaries, in order that need privacy protections, while other links of less privacy concerns are properly released to maintain the graph utility. The goal of TPP is to limit the target disclosure by deleting a budget limited set of alternative non-target links referred to as \emph{protectors} to defend the adversarial link predictions for all targets. Traditional link privacy preserving treated all links as targets and concentrated on structural level protections in which serious link disclosure and high graph utility loss is still the bottleneck of graph releasing today, while TPP focuses on the target level protections in which key protection is implemented on a tiny fraction of critical targets to achieve better privacy protection and lower graph utility loss. Currently there is a lack of clear TPP problem definition, provable optimal or near optimal protector selection algorithms and scalable implementations on large-scale social graphs.

Firstly, we introduce the TPP model and propose a dissimilarity function used for measuring the defense ability against privacy analyzing for the targets. We consider two different problems by budget assignment settings: 1) we protect all targets and to optimize the dissimilarity of all targets with a single budget; 2) besides the protections of all targets, we also care about the protection of each target by assigning a local budget to every target. Moreover, we propose two local protector selections, namely cross-target and with-target pickings. Each problem with each protector picking selection is corresponding to a greedy algorithm. We also implement scalable implementations for all greedy algorithms by limiting the selection scale of protectors, and we prove that all greedy-based algorithms achieve approximation by holding the monotonicity and submodularity. Through experiments on large real social graphs, we demonstrate the effectiveness and efficiency of the proposed target link protection methods.

\end{abstract}

\IEEEoverridecommandlockouts

\IEEEpeerreviewmaketitle


\section{Introduction}
Links of a social graph are rich of private contact behaviors (\emph{e.g.} private friendship and confidential financial transactions between two users), and the link disclosure heavily imperils the individual privacy concerns such as the user deanonymization \cite{ref5}. In this sense, link anonymization is more fundamental than user anonymization, which is the focus of this work. Traditional link privacy preserving technologies \cite{ref1,ref2,ref3,ref4,ref5,ref29,ref30,ref31,ref32,ref33} such as structure perturbations, graph generations and differential privacy provide structural level protections in which all links are treated to be sensitive and extensively perturbed to defend the adversarial attacks. Unfortunately, it is very difficult in protecting all links, and serious link disclosure is still the bottleneck of graph structure releasing today.

In practice, only a limited number of links are significantly important. In a social graph, individuals often like to share most of their social links with other individuals and intentionally hide several private links such as a link that as a patient a person visited a cancer doctor. The disclosure of these important links often leads to serious security disasters (\emph{e.g.} privacy disclosure, economic loss or even life threats). If the link that the patient visited the cancer doctor is exposed, the attackers will infer that the patient got cancer which is the very sensitive privacy for the patient. For another instance, a terrorist might extensively analyze the social links to kidnap one or more important hostages (\emph{e.g.} the parents, spouse, sons/daughters, and important friends or cooperators) to directly threaten a victim. Thus, it is badly in need of protections of these important but vulnerable links (``\emph{targets}'') against the advanced link privacy analyzing methods, and for security purpose, it is not surprising that lots of users hide their important and sensitive relationship links in social graphs such as Facebook and Wechat. However, although the target links are eliminated before graph structure releasing, attackers have the ability of remarkably inferring the missing targets by analyzing the building principles of the graph. Then some other non-target links referred to as \emph{protectors} should be properly deleted to help targets hide better. Then our work is that instead of protecting all links we choose to intensively provide privacy protections for these most important and sensitive targets to satisfy the urgent needs of social graphs.

We model the above scenario by the Target Privacy Preserving (TPP) task. A social graph can be defined as $G=(V,E)$, where $V$ is the node set and $E$ is the edge set. $T$ is defined as the target set, which is a subset of $E$. TPP can be accomplished by two phases. In phase-1, all targets are eliminated to hide themselves first, namely $E=E\setminus T$. In phase-2, an increase dissimilarity function $f(P,T)$ is introduced to quantify the defense ability of TPP against target attacks, where $P (P\subseteq E)$ is the set of protectors which are efficiently selected to optimize the dissimilarity function. The goal of TPP is with a limited budget $k$ which is the maximum deletion number of links, to optimally select protectors for set $P$ ($|P| \le k$) which maximally promotes the dissimilarity scores for all targets.

TPP is a new privacy model and differs from the traditional link privacy preserving, which focuses on target level privacy protections for key targets to satisfy more practical privacy needs of current social graphs.

Budget is a critical constrain of dissimilarity function, and we study two scenarios for budget assignments. (1) All targets share a global budget. Every protector is iteratively selected as the one that increases the dissimilarity of all targets most. We design a greedy algorithm to achieve an $1-1/e$ approximation of optimal solution. (2) Every target $t$ is assigned a sub budget $k_t$ which is mainly used to protect itself and additionally help other targets. To this end, we design two kinds of general protector selections. The first cross-target greedy algorithm globally selects protector cross different targets and achieves an approximation ratio $1/2$. The second within-target greedy algorithm selects protectors target by target and achieve an approximation ratio 0.46.

We further do scalable implementations for all greedy algorithms to run in large-scale social graphs. We conduct experiments on many real social graphs to demonstrate the effectiveness of our methods. We compare the similarity score evolutions for all link deletion methods with two related baselines. With limited deletion budget, the global budget based greedy algorithm can achieve the best protections. We further analyze the graph utility loss to demonstrate the effectiveness of TPP solutions.

The contributions of this work are: (a) mathematically defining a new TPP problem and a dissimilarity function; (b) theoretically proving that the optimal protector selection is NP-hard for TPP and the objective functions are monotone and submodular; (c) proposing three greedy algorithms for three different protector selection scenarios under two budget assignment settings; (d) scalable implementations for all greedy algorithms to run on large-scale social graphs; (e) conducting experiments on many real social graphs to demonstrate the effectiveness and efficiency of proposed algorithms.
\section{Related work}
Many related link privacy preserving technologies have been proposed to limit the link disclosures. Structure perturbations \cite{ref1,ref2,ref3,ref4,ref5} mainly employed randomization algorithms to rewire/switch the real links into fake ones to cheat or defend adversarial link predictions \cite{ref4,ref5,ref8,ref9,ref17}. Graph generation methods \cite{ref11,ref12,ref13,ref14,ref15,ref16} sampled many important graph characteristics (\emph{e.g.} degree distributions and degree correlations) based on which a serial of pseudo graphs was generated to represent the original one. Thus, the structure privacy can be to some extend preserved. Differential privacy mechanisms \cite{ref7,ref10,ref30,ref31,ref32} provided the queries of edge, node, or subgraphs to satisfy $\epsilon$-differential privacy in which given the maximum background knowledge an attacker can't infer the existence of a given edge, node or subgraph respectively. Furthermore, the global \cite{ref33} and local \cite{ref34} differential privacy based graph generations can yield pseudo graphs for privacy protection purpose. Others such as k-means clustering \cite{ref26}, k-isomorphism \cite{ref28,ref29}, k-anonymity \cite{ref27} and L-opacity \cite{XueKRKP12} can also protect subgraph related structure privacy. However, these methods didn't consider target-level privacy protections.

Related link deletion method for privacy preserving is rare, which is often a step for link switching methods \cite{ref2,NobariKPB14}.



\section{Privacy Model and Problem Definition}
Talbe \ref{tab0} lists the main mathematical notations of this work.
\begin{table}[htb]
\newcommand{\tabincell}[2]{\begin{tabular}{@{}#1@{}}#2\end{tabular}}
\caption{The main mathematical notations used in this work.}
\begin{tabular}{ll}
\hline
Symbol & Definition \\
\hline
$G$ & The social \textbf{g}raph \\
V; E & The set of \textbf{v}ertices/nodes; The set of \textbf{e}dges/links \\
 $T$ & The set of \textbf{t}arget links \\
 $P$ & The set of \textbf{p}rotectors \\
 $t$ & A \textbf{t}arget \\
 $p$ & A \textbf{p}rotector \\
 $P_t$ & The set of protectors for target $t$\\
$s(P,t)$ & Similarity of $t$ when deleting $P$ protectors \\
 $s(P,T)$ & Total similarity of $T$ targets when deleting $P$ protectors \\
 $f(P,T)$ & Total dissimilarity of $T$ targets when deleting $P$ protectors \\
 $C$ & A constant number for the dissimilarity function \\
 $k$ & The link deletion budget \\
 $k^*$ & The critical budget that provides full privacy protection \\
 $k_t$ & The sub budget for target $t$ \\
 $K$ & The sub budget vector for all targets \\
 $W_t$ & The set of target subgraphs for target $t$ \\
 $W$ & The set of all target subgraphs for all targets \\
 $\Delta p$ & The dissimilarity gain if deleting $p$ in the SGB-Greedy \\
 $\Delta _p^t$ & \tabincell{l}{Dissimilarity gain if deleting $p$ for $t$ in the WT-Greedy \\and CT-Greedy algorithms} \\
\hline
\end{tabular}
\label{tab0}
\end{table}
\subsection{Target Privacy Preserving Model}
\begin{figure}[ht!] 
\centering
\includegraphics[width=3.4in]{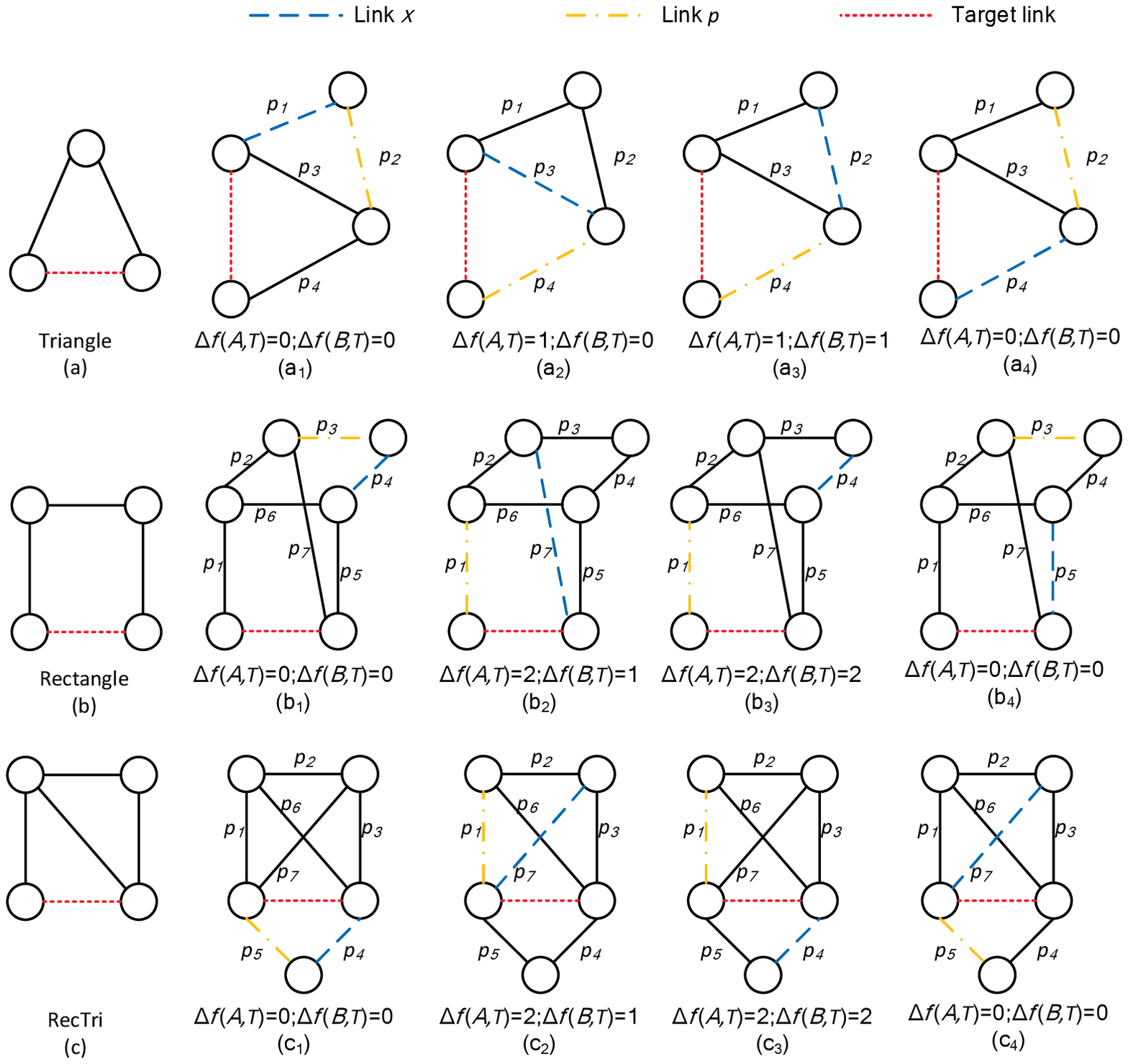}
\caption{The illustrations and examples for the proof of submodularity for the dissimilarity function by using the three classical subgraph patterns: \emph{Triangle}, \emph{Rectangle}, and \emph{RecTri}.
}
\label{fig1}
\end{figure}

The design of TPP is to efficiently defend some specific adversarial link predictions. In this work, we mainly focused on the subgraph pattern or motif \cite{ref40} based link prediction. In general, a missing link is predictable attributing to its frequent participation into a specific \emph{subgraph pattern} or \emph{motif} \cite{ref40} which is the building principle of most real social graphs. Based on the principle, adversaries can infer the missing targets and disclose the privacy. The \emph{Triangle} motif as shown in Fig.~\ref{fig1}(a) has been widely used for link predictions. If two ends of a missing link can communicate to each other via at least one 2-length paths, the existing probability for the missing link is proportional to the number of 2-length paths between the two ends. It is the basis of common neighbor related link predictions. To extend, if the two ends of a missing link are routed by multiple 3-length paths, we can also infer that there is an existing probability for the missing link. For instance, in a social graph, if two users are initially not friends, but the friends of the two users are strongly connected, there exists a high probability of building friendship introduced by the friends of friends. This case is based on the \emph{Rectangle} motif as shown in Fig.~\ref{fig1}(b). Furthermore, the missing link might frequently participate into some complex patterns, for instance in Fig.~\ref{fig1}(c). In this pattern, the two users are simultaneously and indirectly connected by a 2-length path and a 3-length path which shares an intermediate node with the 2-length path, and we treat this pattern as \emph{RecTri} motif which can be considered as a classical representation of complex patterns. In fact, it is general to use any motif as link prediction basis in TPP. Without loss of generality and for simplicity, in this work we use \emph{Triangle}, \emph{Rectangle} and \emph{RecTri} as three motif instances.

A subgraph is regarded as a \emph{target subgraph} denoted by $w_t$ for a target $t$ together with which it is in the form of focused motif such as \emph{Triangle}, \emph{Rectangle}, and \emph{RecTri} in Fig.~\ref{fig1}. For a specific subgraph pattern, we denote all target subgraphs for target $t$ by $W_t$ in the graph. Because all targets have been removed in phase-1, in phase-2 a target subgraph can be only included in one target subgraph set, namely for any two target subgraph sets $W_t$ and $W_{t'}$, $W_t \cap W_{t'} = \emptyset$. The number of target subgraphs for a target $t$ is generally referred to as \emph{similarity} denoted by $s(P,t)=|W_t|$ for $t$, where a higher similarity of a target link means higher probability being inferred. The total similarity for all targets is defined as $s(P,T)=\sum_{t\in T}s(P,t)$, which is the vulnerability of being attacked for all targets. Meanwhile, we define a dissimilarity function $f(P,T)=C-s(P,T)$ to indicate the attack defense ability of all targets against link predictions, where $C$ is a constant and large number satisfying $C \ge s(\emptyset,T)$ and $f(P,T) \ge 0$, and higher dissimilarity means lower probability of being inferred or attacked for all targets. The $f(P,T)$ is an increase function of monotonicity and submodularity. A protector might participate in multiple target subgraphs, for instance protector $p_3$ in Fig.~\ref{fig2}(a) included in two target subgraphs. Deleting the protector (\emph{e.g.} $p_3$) can increase the dissimilarity score by equal number of the broken target subgraphs (\emph{e.g.} 2 for deleting $p_3$). The task of TPP is to optimally select protectors for set $P (|P| \le k)$ which maximizes the dissimilarity function $f(P,T)$.

The $f(P,T)$ is strongly limited by the budget, namely $|P| \le k$. We consider two budget-related TPP problems in the following sub Sec.~\ref{secc} and \ref{secd}.
\subsection{Threat Model}
We assume the adversarial attackers have the full knowledge of the privacy-preserved graph based on which the attackers predict the existence of these hidden links (\emph{i.e.} targets) by employing a specific link prediction method (\emph{e.g.} subgraph-based link prediction) or link inference (\emph{e.g.} Bayesian inference).

\subsection{Single-Global-Budget based TPP Problem (SGBT)}
\label{secc}
All targets share a single global budget $k$, and up to $k$ protectors are globally selected and eliminated through up to $k$ iterations. We aim to find a such protector set $P$ that can achieve the maximal dissimilarity scores.

\emph{Definition} 1. The TPP under Single-Global-Budget condition is the optimization task where inputs include the graph $G=(V,E)$, the target set $T$, and budget $k$. The goal is to find a protector set $P$ ($|P| \le k$) such that the dissimilarity $f(P,T)=C-\sum_{t \in T} s(P,t)$ is maximized.

\emph{Theorem 1}. The optimal protector set selection for SGBT problem is NP-hard.

\emph{Proof} (Sketch). All target subgraphs for all targets is denoted by $W=\bigcup_{t \in T} W_t$. A protector $p$ might participate in multiple target subgraphs. Deleting the protector $p$ ($p \in E$) might break a set $Q_p$ ($Q_p \subseteq W$) of target subgraphs. To maximize the dissimilarity of all targets is to maximally break the target subgraphs with a limited deletion budget $k$. Assuming every link in $E$ can be considered as one protector, and we have the set $Q=\{Q_p |p \in E\}$.  Our aim is converted to find a subfamily $P$ ($|P| \le k, P \subseteq E$) to achieve max($|\bigcup_{p \in P} Q_p|$). It is the traditional \emph{Max-k-Cover} problem \cite{ref41} for which the optimal solution is NP-hard.

\subsection{Multi-Local-Budget based TPP Problem (MLBT)}
\label{secd}
Every target $t$ is assigned a sub budget $k_t$ ($0 \le k_t \le k$), and $K=\{k_t |t \in T\}$ is the set of all sub budgets. The budget $k_t$ is used for preserving the privacy of target $t$.

\emph{Definition} 2. The TPP under Multi-Local-Budget condition is the optimization task where inputs include the graph $G=(V,E)$, the target set $T$, and sub budget set $K$. The goal is to find $|T|$ sub protector sets of which each set $P_t$ ($|P_t | \le k_t$) for target $t$ and the set $P=\bigcup_{t \in T} P_t$, such that the dissimilarity function $f(P,T)=C-\sum_{t \in T} s(P_t,T)$ is maximized.

\emph{Theorem} 2. The optimal protector set selection for MLBT problem is NP-hard.

\emph{Proof} (Sketch). Like the \emph{Theorem} 1, every deleted protector $p$ for any target can break a set $Q_p$ ($Q_p \subseteq W$) of target subgraphs. The goal is to find the \emph{Max-k-Covers} \cite{ref41}.  Then the optimal protector selection for MLBT is also NP-hard.

\section{Optimal solution for the SGBT problem}
After deleting all targets from the edge list of original graph, one specific target subgraph such as \emph{Triangle} can only include one target. Every target subgraph in $W$ only serves for one target. Meanwhile, every removable protector might serve for multiple targets, or a protector might participate in multiple target subgraphs for one target.

\emph{Lemma} 1. The dissimilarity function $f(P,T)$ of SGBT is monotone, for any set relation $A \subseteq B \subseteq P$, $f(A,T) \le f(B,T)$.

\emph{Proof}. By \emph{Definition} 1,

$f(A,T)-f(B,T)=\sum_{t \in T} s(B,t) - \sum_{t \in T}s(A,t) = \sum_{t \in T}(s(B,t)-s(A,t))$

Without loss of generality, we assume $B=A \cup \{p\}$, then

$f(A,T)-f(B,T) = \sum_{t \in T}(s(A\cup \{p\}, t)-s(A,t))$

For any target $t$, there are at most two cases for a new deleted protector $p$.

\emph{Case} 1. Protector $p$ is included in at least one target subgraph of $W_t$. In this case, one or more subgraphs will be broken from the graph, and the number of target subgraphs for target $t$ decreases, namely $s(A \cup \{p\},t)-s(A,t)<0$.

\emph{Case} 2. Protector $p$ is not included in any target subgraph of $W_t$, then $s(A \cup \{p\},t)-s(A,t)=0$.

Then for any target $t$, the $s(A\cup \{p\},t)-s(A,t) \le 0$ can be guaranteed, so we can infer that $\sum_{t \in T}(s(A \cup \{p\},t)-s(A,t)) \le 0$, namely $f(A,T)-f(B,T) \le 0$. The monotonicity is proved.

\emph{Lemma} 2. The dissimilarity function $f(P,T)$ of SGBT is submodular, for the sets $A\subseteq B\subseteq P$, $f(A\cup\{p\},T)-f(A,T)\ge f(B\cup\{p\},T)-f(B,T)$, where $p \in E\setminus B$.

\emph{Proof}. By \emph{Definition} 1

$\Delta f(A,T) = f(A\cup \{p\},T) - f(A,T) = \sum_{t\in T} (s(A,t)-s(A\cup \{p\},t))$

Similarly, $\Delta f(B,T)=f(B\cup \{p\},T)-f(B,T) = \sum_{t\in T}(s(B,t)-s(B\cup \{p\},t))$

In fact, $\Delta f(A,T)$ and $\Delta f(B,T)$ are the reduced number of target subgraphs if a protector $p$ is deleted from graph $G=(V,E\setminus A)$ and $G=(V,E \setminus B)$ respectively.

Without loss of generality, here we assume $B=A\cup\{x\}$, where $x$ is a deleted protector in $B$ but not in $A$. For any target $t$, there are at most four cases for the location combinations of protector $p$ and $x$ in any target subgraph pattern as shown in Fig.~\ref{fig1}.

\emph{Case} 1. Both protector $p$ and $x$ are not the edges of any target subgraph for target $t$, for instance in Fig.~\ref{fig1}($a_1$) for \emph{Triangle} subgraph $p=p_2$ and $x=p_1$, then $\Delta f(A,t)=s(A,t)-s(A\cup \{p\},t)=0$ and $\Delta f(B,t)=s(B,t)-s(B\cup \{p\},t)=0$. $\Delta f(A,t) = \Delta f(B,t)=0$.

\emph{Case} 2. Both $p$ and $x$ are the edges of any given target subgraph for target $t$, still taking \emph{Triangle} for example in Fig.~\ref{fig1}($a_2$) $x=p_3$, $p=p_4$. For set $A$, link $x$ is still in its target subgraph, deleting protector $p$ will lead to $\Delta f(A,t)=1$. Meanwhile, if link $x$ is deleted in advance in set $B$, deleting protector $p$ will lead to $\Delta f(B,t)=0$. It can be seen that $\Delta f(A,t) > \Delta f(B,t)$.

\emph{Case} 3. Protector $p$ included in at least one target subgraph of target $t$, and $x$ beyond any target subgraph of target $t$, for instance in Fig.~\ref{fig1}($a_3$), $x=p_2$ and $p=p_4$, then we have $\Delta f(A,t)= \Delta f(B,t)=1$.

\emph{Case} 4. Link $p$ and $x$ are beyond and in any target subgraph of target $t$ respectively. As in Fig.~\ref{fig1}($a_4$) for \emph{Triangle} pattern, we randomly set $x=p_4$ and $p=p_2$, and then we have $\Delta f(A,t)=  \Delta f(B,t)=0$.

Thus, we can see that for all the four cases discussed above, for any target $t \in T$, the equation $\Delta f(A,t) \ge \Delta f(B,t)$ can be absolutely guaranteed. Then we have $\Delta f(A,T)=\sum_{t \in T} \Delta f(A,t)$  , $\Delta f(B,T) = \sum_{t\in T} \Delta f(B,t)$, and it can be inferred that $\Delta f(A,T) \ge \Delta f(B,T)$. The submodularity property of the dissimilarity function $f(P,T)$ is proved.

\emph{Theorem} 3. The SGBT can yield an $1-1/e$ approximation of the optimal solution.

\emph{Proof} (Sketch). SGBT is converted to the classical \emph{Set Cover} \cite{ref41} problem which has monotonicity and submodularity property. It has been proved that the greedy solution for set cover has an $1-1/e$ approximation of optimal solution.

Then the SGBT problem can be solved by employing a greedy algorithm to achieve a near optimal solution. At every step, the number of broken target subgraphs for every alternative link in the graph is computed, and we select the link as a protector which can break the highest number of target subgraphs for all targets. The process can be described by Algorithm \ref{alg1}.

\begin{algorithm}
\caption{SGB-Greedy: The Single-Global-Budget Greedy Protector Selection Algorithm}
\label{alg1}
\KwIn{Graph $G=(V,E)$, target set $T$, and budget $k$.}
\KwOut{The protector set $P$}
$P=\emptyset$ \;
\While{$|P|<k$}{
\ForEach{$p \in E$}{
estimate $\Delta p =f(P \cup \{p\},T)-f(P,T)$ \;
}
find $p^* \leftarrow argmax_{p \in E} \Delta p$ \;
\If{$\Delta p^*==0$}{
return\;
}
$P=P\cup \{p^*\}$ \;
$E=E\setminus \{p^*\}$;
}
\end{algorithm}

In the Algorithms \ref{alg1}, for each selected protector, the dissimilarity gains of all alternative links are recalculated. The time complexity for calculating the similarity score $s(P,t)$ is to search the number of target subgraphs it participates in. For different subgraph patterns, the time complexity is different. For the motif instances used in this work, the similarity calculating time complexity for any target $t=(u,v)$ is $O(d_u d_v)$ where $d_u$ and $d_v$ are the degrees of node $u$ and $v$ respectively. In general, the degree of a node is proportional to $\log N$ where $N$ is the networks size \cite{ref13}. The time complexity for calculating the dissimilarity score is $O(n (\log N)^2)$ where $n$ is the number of targets. At each step, every link is tried as a protector, and the time complexity is $O(mn(\log N)^2)$. Total time complexity for selecting at most $k$ protectors is $O(kmn (\log N)^2)$.

\section{Optimal solution for the MLBT problem}
In real application settings, relationship strengths between all pairs of nodes are heterogeneous, and the importance level of every sensitive target is different. For a graph releaser, with a finite budget $k$, it is critical to primarily protect the privacy of more important targets and assign higher budget for them.
In general, with a total budget $k$, based on a specific assignment strategy, every target $t$ is assigned a sub budget $k_t(\sum_{t\in T}k_t \le k)$. Budget $k_t$ is mainly used for the target $t$. The set $P_t$ is a subset of set $P$ and contains the alternative protectors that can significantly reduce the target subgraphs in $W_t$. A protector in $P_t$ also can help other targets break their target subgraphs. Then for multi-local-budget TPP problem, it still needs global graph structure to globally pick alternative protectors for all targets.

\emph{Lemma} 3. For MLBT problem in \emph{Definition} 2, the dissimilarity function $f(P,T)=C- \sum_{t \in T} s(P_t,T)$ is monotone, For any $t$, $A_t \subseteq B_t \subseteq P_t$, $A=\bigcup_{t\in T}A_t$, $B=\bigcup_{t\in T} B_t$, $f(A,T) \le f(B,T)$.

\emph{Proof}. By \emph{Definition} 2

$f(A,T)-f(B,T)=\sum_{t\in T}(s(B_t,T)-s(A_t,T))$

By \emph{Lemma} 1, for any target $t$, the sets satisfy $A_t \subseteq B_t \subseteq P_t$, we have
$s(B_t,T)-s(A_t,T) \le 0$

Then for all targets, we have $\sum_{t \in T} (s(B_t,T)-s(A_t,T)) \le 0$. Thus, the monotonicity property is proved.

\emph{Lemma} 4. For MLBT problem in \emph{Definition} 2, the dissimilarity function $f(P,T)=C-\sum_{t\in T}s(P_t,T)$ is submodular. For any target $t$, $A_t\subseteq B_t\subseteq P_t$, $A=\bigcup_{t\in T}A_t$, $B=\bigcup_{t\in T}B_t$, $f(A\cup \{p\},T)-f(A,T) \ge f(B\cup \{p\},T)-f(B,T)$.

\emph{Proof}. By \emph{Definition} 2

$\Delta f(A,T)=f(A\cup \{p\},T)-f(A,T)
=\sum_{t\in T}(s(A_t,T)-s(A_t\cup \{p\},T))$

Similarly, $\Delta f(B,T)=\sum_{t\in T}(s(B_t,T)-s(B_t\cup \{p\},T))$

By \emph{Lemma} 2, for any target $t$, if $A_t\subseteq B_t \subseteq P_t$, we have

$s(A_t,T)-s(A_t\cup \{p\},T) \ge s(B_t,T)-s(B_t\cup \{p\},T)$

Then for all targets, we have $\Delta f(A,T) \ge \Delta f(B,T)$. Therefore, the submodularity is proved.
For the problem in \emph{Definition} 2, there are two aspects to be properly solved. 1) Budget division, where the sub budget for every target is given. 2) How to select protectors for every target to achieve the most dissimilarity increase?

\subsection{Budget Division Strategies}
We consider two categories of budget division methods. Firstly, for any target $t$, $k_t \le |W_t |$ is constricted. A target of more target subgraphs needs more budget to protect its privacy. The sub budget $k_t$ for a target $t$ can be designed proportionally to the number of target subgraphs for target $t$, and it is denoted as the target-subgraph-based budget division (TBD).

Every link is often related to its two ends. The degree product of the two ends can also be considered as the metric of link importance in the network. Assuming the degree of the two ends (\emph{i.e.} $u$ and $v$) of any target $t$ as $d_u$ and $d_v$ respectively, the budget $k_t$ for any target $t$ can be defined proportional to the degree product of the two ends. It is called as the degree-product-based budget division (DBD) strategy.

For MLBT problem, when selecting an alternative protector to remove, there are two considerations: 1) local budget $k_t$ is mainly used for decreasing the target subgraphs for the target $t$, and 2) it may also help other targets to break their target subgraphs. Given a local budget division $K$, there are two settings for protector picking: Cross-Target setting and Within-Target setting.
\subsection{Cross-Target Protector Selection for MLBT}
For cross-target protector selection, given a set $P$ which is partitioned into disjoint sets $P_1$,$P_2$,$\dots$,$P_n$ and $I=\{X\subseteq P:|X \cap P_t| \le k_t,\forall t\in T\}$, $(P,I)$ is called a partition matroid \cite{ref35,ref36}. We use pair $(t,p)$ to denote deleting a link $p$ for target $t$. For every target and every protector, the scores are calculated, and we select the one whose deletion increase the dissimilarity score most. The protector selection crosses all the targets whose budgets are not used up. $\Delta_p^t$ represents the total increased dissimilarity score by deleting a protector $p$ for target $t$.  $\Delta_p^t$ includes two parts of dissimilarity increasement: 1) $s(P_t,t)-s(P_t\cup \{p\},t)$ indicates the number of broken target subgraphs in the $W_t$ for this target; 2) $\sum_{t'\in T\setminus \{t\}}(s(P_t,t')-s(P_t \cup \{p\},t'))$ is the number of broken target subgraphs for other targets. Such case may exist that $\Delta_p^t=1+4$ where the protector $p$ breaks 1 target subgraph in $W_t$, and 4 in $W\setminus W_t$, while $\Delta_p^t=2+2$ where the protector $p'$ breaks 2 target subgraph in $W_t$, and 2 in $W\setminus W_t$.  Although the protector $p$ breaks more total target subgraphs than $p'$, but $p'$ breaks more target subgraphs for the target $t$ and should be chosen. It is described by Algorithm \ref{alg2}.

\begin{algorithm}
\caption{\emph{CT-Greedy}: The Cross-Target Greedy Protector Selection Algorithm}
\label{alg2}
\KwIn{Graph $G=(V,E)$, target set $T$, budget $k$, $K=TBD(k,W)$ or $DBD(k,W,ds)$.}
\KwOut{The protector set $P$}
$P=\emptyset$ \;
\ForEach{$p \in E$}{$P_t=\emptyset$ \;} 
$T' \leftarrow \emptyset$; //the set of targets whose budgets are used up \
\ForEach{$t \in T$}{
\ForEach{$t \in T\setminus T'$, $p\in E$}{
estimate $\Delta _p^t=s(P_t,t)-s(P_t\cup \{p\},t) + \sum_{t'\in T\setminus \{t\}} [s(P_t,t' )-s(P_t \cup \{p\},t' )]/C$ \;
}
$(t^*,p^*) \leftarrow argmax_{t\in T\setminus T', p \in E} \Delta_p^t$ \;
\If{$\Delta_{p^*}^{t^*}==0$}{
return\;
}
\If{$|P_{t^*}| \ge k_{t^*}$}{$T' \leftarrow T' \cup \{t^*\}$ \;}
$P_{t^*} = P_{t^*} \cup \{p^*\};$ \;
$E=E\setminus \{p^*\}$;
}
\end{algorithm}

For the CT-Greedy algorithm, for each $(t,p)$, all links in $E$ are used, and the time complexity is also $O(mn (\log N)^2)$. The total time complexity is $O(knm (\log N)^2)$ too.

\emph{Theorem} 4. The CT-Greedy algorithm can achieve an approximation $1/2$ of the optimal solution.

\emph{Proof} (Sketch). As discussed above, the cross-target setting is an instance of submodular maximization of partition matroid, and the performance of CT-Greedy satisfies the guarantee \cite{ref35,ref36}. Then the final output of CT-Greedy can achieve at least $1/2$ approximation of the optimal solution.
\subsection{Within-Target Protector Selection for MLBT}
For within-target scenario, we can greedily pick protectors for the first target. After the first target satisfied, the second target is greedily satisfied, until sub budgets of all targets are used up. The process can be described by Algorithm \ref{alg3}.

\begin{algorithm}
\caption{\emph{WT-Greedy}: The Within-Target Greedy Protector Selection Algorithm}
\label{alg3}
\KwIn{Graph $G=(V,E)$, target set $T$, budget $k$, $K=TBD(k,W)$ or $DBD(k,W,ds)$.}
\KwOut{The protector set $P$}
$P=\emptyset$ \;
\ForEach{$\forall p \in E$}{$P_t=\emptyset$ \;}
\ForEach{$t \in T$}{
\For{$b=1,2,3,\cdots, k_t$}{
\ForEach{$t \in T\setminus T'$, $p \in E$}{
estimate $\Delta _p^t=s(P_t,t)-s(P_t\cup \{p\},t) + \sum_{t'\in T\setminus \{t\}} [s(P_t,t' )-s(P_t \cup \{p\},t' )]/C$ \;
}
$(t,p^*) \leftarrow argmax_{t\in T\setminus T', p \in E} \Delta_p^t$ \;
\If{$\Delta_{p^*}^t==0$}{
return\;
}
$P_{t} = P_{t} \cup \{p^*\};$ \;
$E=E\setminus \{p^*\}$;
}
}
\end{algorithm}

In Algorithm \ref{alg3}, within each target $t$ of sub budget $k_t$, the time complexity is $O(k_t mn(\log N)^2)$. For all targets, the total time complexity is $O(mn (\log N)^2 \sum_{t \in T}k_t)$.

\emph{Theorem} 5. The WT-Greedy algorithm can achieve an approximation $1-e^{-(1-1/e)}$ of the optimal solution.

\emph{Proof} (Sketch). The within-target setting is also an instance of submodular maximization for picking the protectors restricted by the local budget for every target link. It has been extensively discussed in \cite{ref35,ref36}, and the bound has been proved to be $1-e^{-(1-1/e)}$. Therefore, the WT-Greedy in our work can achieve an approximation ratio $1-e^{-(1-1/e)}$.

\begin{figure}[ht!] 
\includegraphics[width=3.1in]{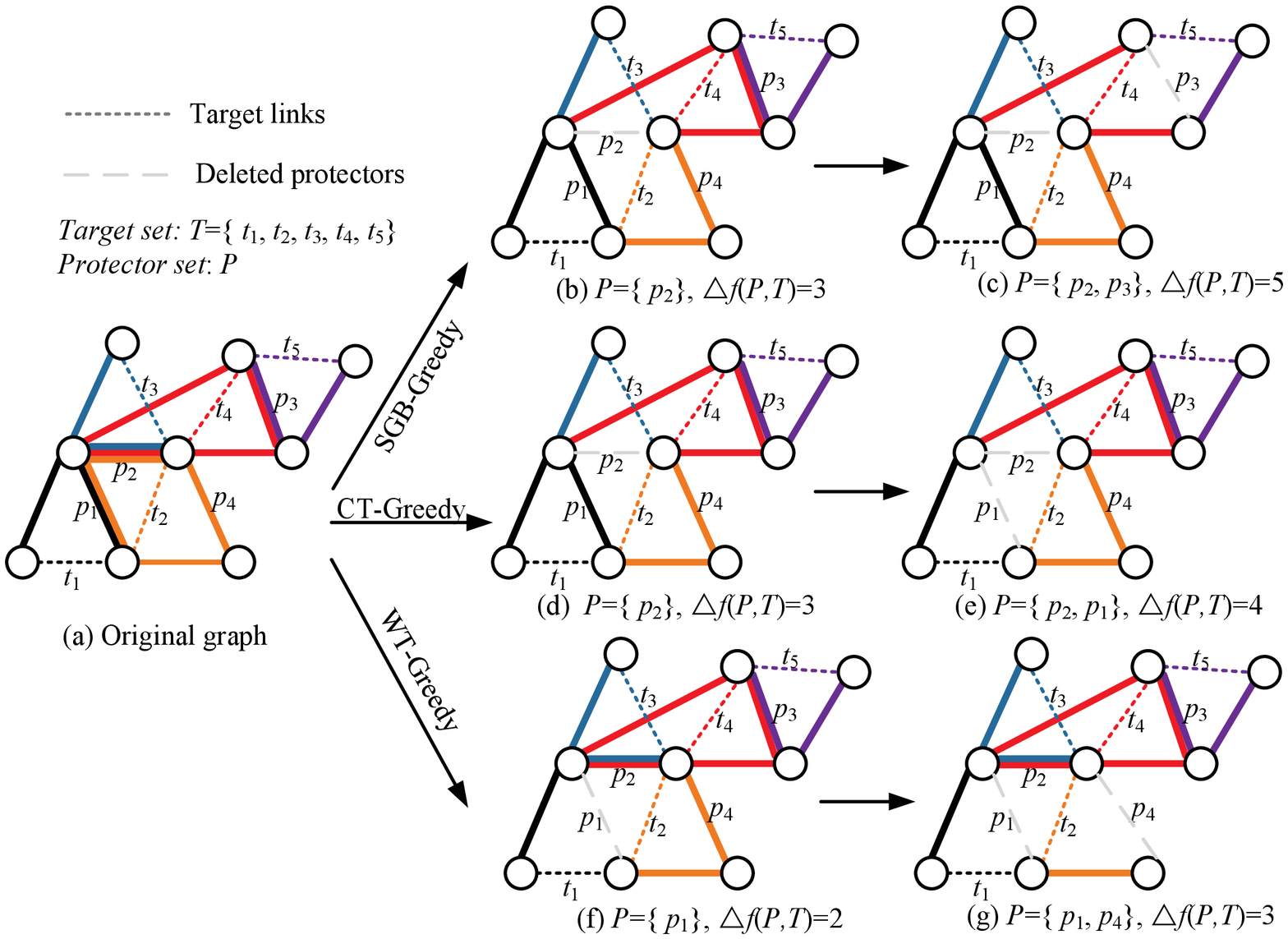}
\caption{An example to illustrate the difference among the SGB-Greedy, CT-Greedy, and WT-Greedy algorithms.
}
\label{fig2}
\end{figure}

The SGB-Greedy globally uses the budget. The CT-Greedy can select every protector globally cross all targets for the target who still has budget. The WT-Greedy only pick protectors within a given target. For these three settings, in Fig.~\ref{fig2}, we give an example (using the \emph{Triangle} pattern) to illustrate the comparisons of these methods. In Fig.~\ref{fig2}(a), there are 5 targets in this example, and link $p_1$ participates in 2 target triangles for target $t_1$ and $t_2$; link $p_2$ in 3 target triangles for $t_3$, $t_4$ and $t_2$; link $p_3$ in 2 target triangles for $t_4$ and $t_5$. We assume the total budget is 2 and assign sub budget 1 for $t_1$ and $t_2$ (others 0) under CT-Greedy and WT-Greedy. By SGB-Greedy, two protectors are selected as $p_2$ and $p_3$ in Fig.~\ref{fig2}(b) and Fig.~\ref{fig2}(c) respectively. The increased dissimilarity score is 5. By CT-Greedy, the increased score is 4, by deleting $p_2$ and $p_1$ for target $t_2$ and $t_1$ in Fig.~\ref{fig2}(d) and Fig.~\ref{fig2}(e) respectively. By WT-Greedy, the increased score is 3, deleting $p_1$ and $p_4$ for target $t_1$ and $t_2$ in Fig.~\ref{fig2}(f) and Fig.~\ref{fig2}(g) respectively.
\subsection{Scalable Implementations}
In the above SGB-Greedy, CT-Greedy, and WT-Greedy algorithms, for every protector selection, the increased dissimilarity scores of all alternative links in the network are computed. In real large social graphs such as the Facebook, it is very time-consuming to estimate the dissimilarity scores for all links. In fact, only the links that participate in the target subgraphs can break target sub-graphs.

\emph{Lemma} 5. The alternative protectors are restricted to these links of target subgraphs.

\emph{Proof} (Sketch). It is straightforward, because deleting links beyond the target subgraphs won't change the number of target subgraphs.

It is straightforward to reduce the time consumption of all proposed algorithms by employing \emph{Lemma} 5. The protector selection scale is extensively reduced. The above three greedy algorithms under the restricted condition in \emph{Lemma} 5 are referred to as SGB-Greedy-R, CT-Greedy-R and WT-Greedy-R for the SGB-Greedy, CT-Greedy and WT-Greedy respectively.
\section{Experimental Evaluation}
\subsection{Baselines}
To our best knowledge, our work is the first to study the target privacy preserving problem. There is no available related baseline to use, so we design two related baselines.

1) \emph{Random deletions} (RD). It is the simplest baseline by randomly removing a given number $k$ of links from the edge set $E$.

2) \emph{Random deletions from target subgraphs} (RDT). This baseline is to randomly select $k$ links from many of the target subgraphs.

\subsection{Datasets}
In this work, without loss of generality, we use two widely used datasets of social graphs in our experiments.

\emph{Arenas-email}\footnote{http://konect.uni-koblenz.de/networks/arenas-email}. This is the email communication network at the University Rovira i Virgili in Tarragona in the south of Catalonia in Spain. Nodes are users, and every edge represents that at least one email was sent between the pair of users. Because the direction or the number of emails is not stored, this network is unweighted and undirected, including 1133 nodes and 5451 links.

\emph{DBLP}\footnote{http://snap.stanford.edu/data/com-DBLP.html}. This network is a co-authorship network. Nodes are authors, and each edge represents that the two authors co-authored and published at least one paper together. There are 317080 nodes, and 1049866 edges.

\subsection{Results}
Without loss of generality, the targets are randomly sampled from the existing links of the original graph. We first delete these targets from the edge list, and then we run the SGB-Greedy, WT-Greedy, WT-Greedy and relative scalable algorithms (\emph{i.e.} SGB-Greedy-R, WT-Greedy-R, WT-Greedy-R) separately on a server of 128G RAM. We run every algorithm on at least 10 independent target samplings.

\begin{figure*}[ht!] 
\includegraphics[width=4.35in]{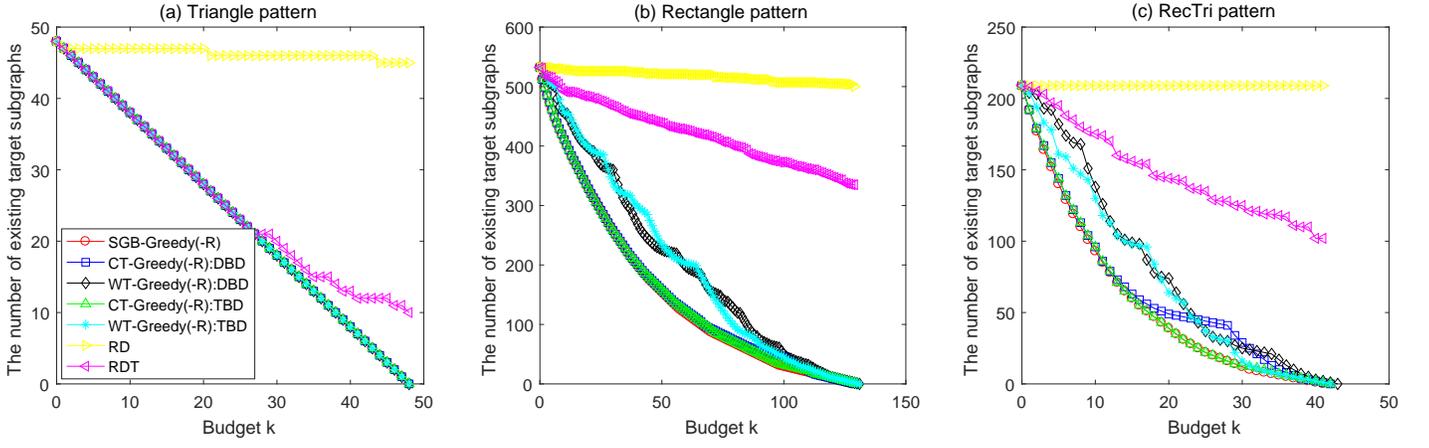}
\caption{Experiments for the evolution of the number of existing target subgraphs as a function of budget $k$ on \emph{Arenas-email} graph.  }
\label{fig3}
\end{figure*}

\emph{Evolution of the number of target subgraphs}. Deleting protectors will cause the increase of dissimilarity scores. In fact, it can be alternatively represented by the similarity which is easy to be observed and understood for result analysis, where lower total similarity means better privacy protection for all targets.
We set the number of targets $|T|=20$, and take budget $k$ as a variable for three subgraph patterns. In Fig.~\ref{fig3}, on the original \emph{Arenas-email} graph, $s(\emptyset,T)$=48, 532, and 209 for the \emph{Triangle}, \emph{Rectangle}, and \emph{RecTri} respectively. Higher $s(\emptyset,T)$ means higher challenge to defend the specific adversarial link predictions. The \emph{Rectangle} motif based TPP seems more challenging.

We set budget $k$ beginning from 1 to the maximum (denoted by $k^*$) that makes $s(P,T)=0$ for every greedy algorithm. As shown in Fig.~\ref{fig3}, with a given budget $k$, for any specific motif, the SGB-Greedy achieves the lowest similarity meaning the highest dissimilarity, because the SGB-Greedy greedily finds every protector that can maximally increase the dissimilarity scores of all targets. With local budget settings, the CT-Greedy method is a bit better than WT-Greedy. For the two budget division strategies, for a specific $k$, the $K$ of TBD leads to lower similarity than that of DBD, because the TBD allocates higher sub budget for a target of higher initial similarity. TBD is more efficient than DBD, but TBD needs to know the initial similarity of every target in advance. RD method randomly select protectors to delete and has the lowest performance. Since the targets are randomly sampled from the existing links, in the \emph{Triangle} pattern case, it is very rare that one protector participates in multiple target triangles. Thus, the RDT seems to achieve very similar similarity score evolution of the three greedy algorithms. For the \emph{Rectangle} and \emph{RecTri} patterns, the RDT has worse performance than the greedy algorithms. Moreover, the $k^*$ for \emph{Rectangle} is the highest, and it confirms that it is more challenging in defending the \emph{Rectangle} motif based adversaries than the other two motifs related ones.

\begin{figure*}[ht!] 
\includegraphics[width=4.35in]{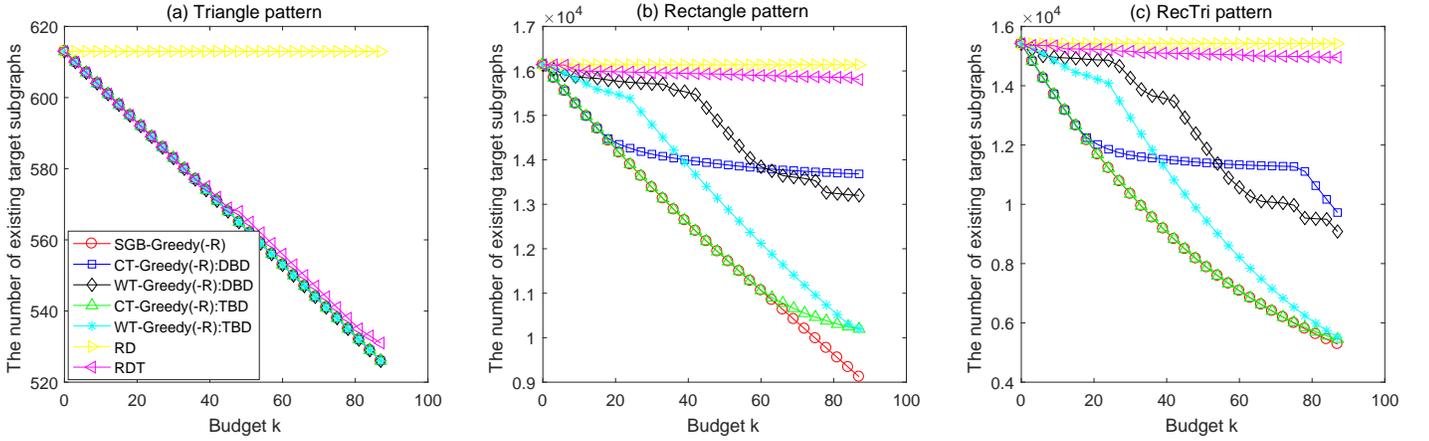}
\caption{Experiments for the evolution of the number of existing target subgraphs as a function of budget $k$ on \emph{DBLP} graph.  }
\label{fig4}
\end{figure*}

Similar results are obtained on \emph{DBLP} graph. The experiments on SGB-Greedy, CT-Greedy and WT-Greedy are extensively long such that every of them didn't finish in one week. Then we show the results under the scalable implementations as shown in Fig.~\ref{fig4}. The similarity under SGB-Greedy-R and CT-Greedy-R: TBD decreases faster than other methods.

It is interesting that for \emph{Triangle} motif subgraph pattern, all methods except for the RD and RDT can achieve very near evolution of similarity score. This is because the number of targets in our experiments is small, and for the triangle subgraph patterns, there is a rare that two target subgraphs share a common link.

\emph{Running time}. It directly reveals the computing efficiency of every algorithm. We compare the running time of all methods on the \emph{Arenas-email} and scalable ones on the \emph{DBLP}.

\begin{figure*}[ht!] 
\includegraphics[width=4.35in]{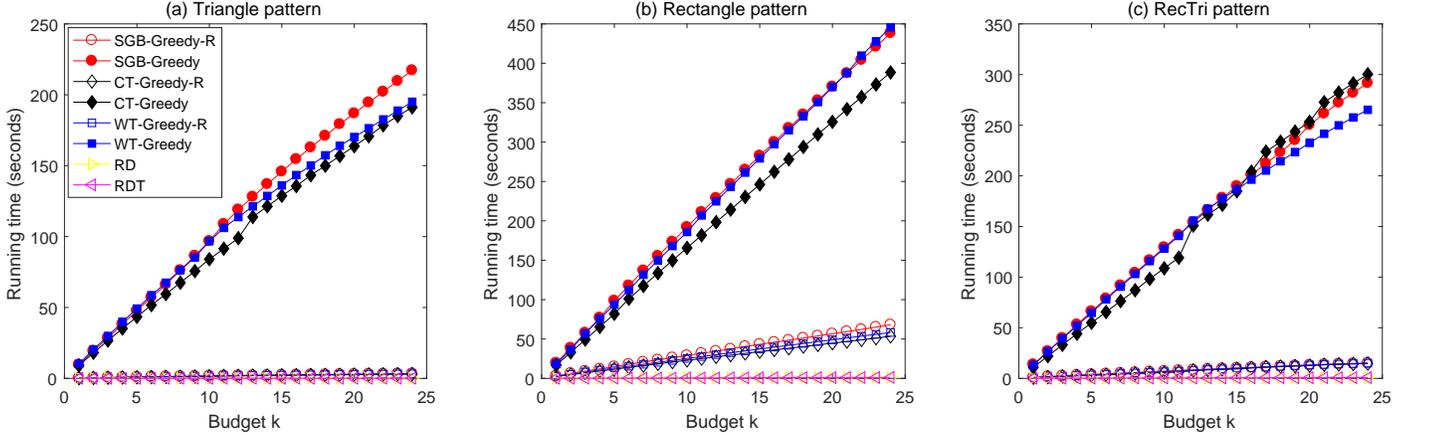}
\caption{Experiments for the evolution of the running time as a function of budget $k$ on \emph{Arenas-email} graph.  }
\label{fig5}
\end{figure*}

As shown in Fig.~\ref{fig5}, for the three subgraph patterns, the running time of all normal greedy algorithms (\emph{i.e.} SGB-Greedy, CT-Greedy, WT-Greedy) is about 20 times more than of the respective scalable implementations (\emph{i.e.} SGB-Greedy-R, CT-Greedy-R, WT-Greedy-R) on the \emph{Arenas-email} graph. For a specific motif, the running time of SGB-Greedy, CT-Greedy and WT-Greedy is very close in accordance with the analyzed time complexity.

\begin{figure*}[ht!] 
\includegraphics[width=4.35in]{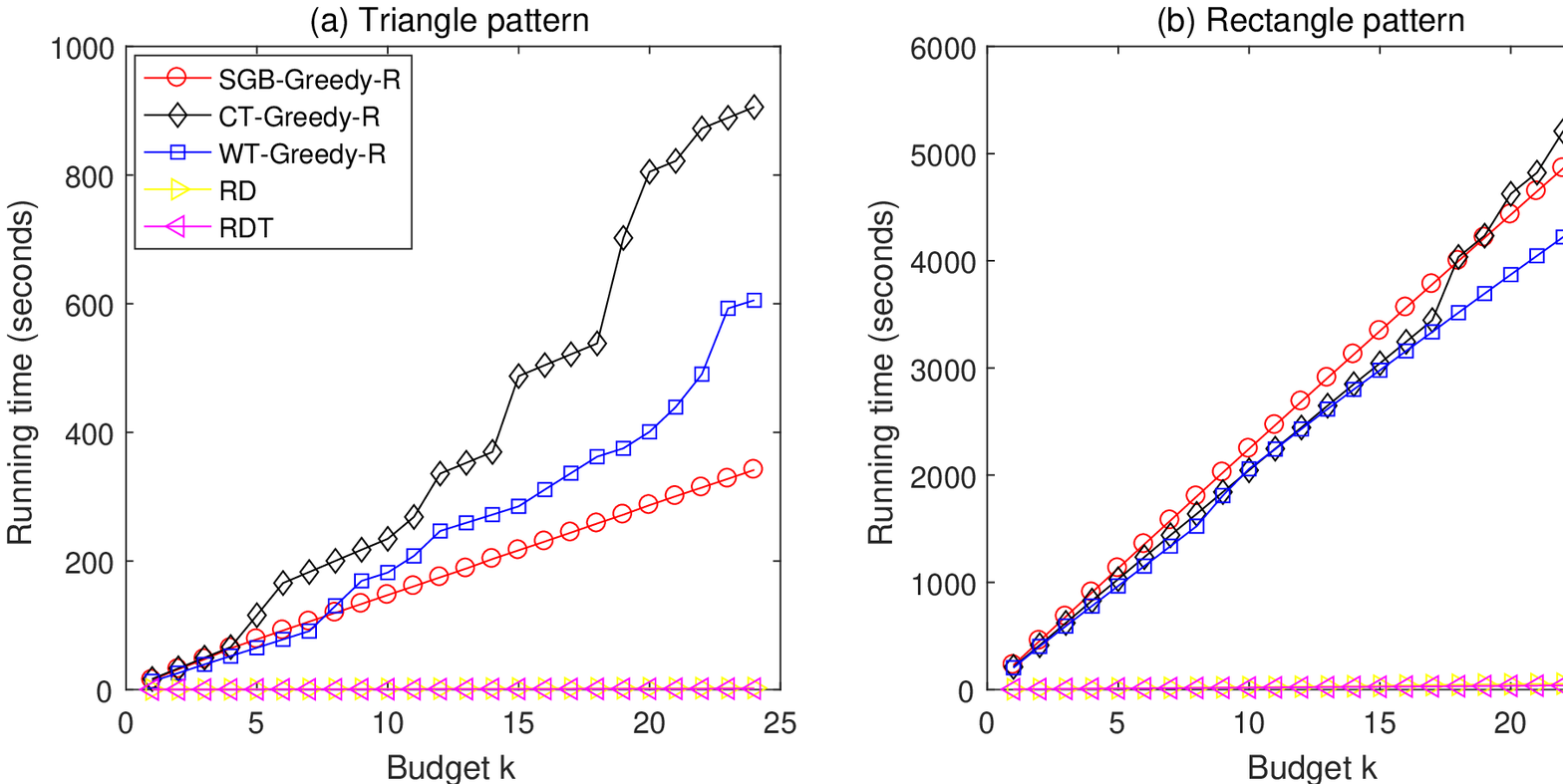}
\caption{Experiments for the evolution of the running time as a function of budget $k$ on \emph{DBLP} graph.  }
\label{fig6}
\end{figure*}

On \emph{DBLP} graph, we set $|T|=50$ and $k=25$. The running time of the three scalable algorithms is within several thousands of seconds for all subgraph patterns used in this work, while the greedy algorithms without scalable implementations didn't finish within a week. As shown in Fig.~\ref{fig6}, we compare the running time for SGB-Greedy-R, CT-Greedy-R, WT-Greedy-R, RD and RDT methods. RD and RDT randomly select protectors from the alternative links without calculating the dissimilarity metric, which has the lowest running time. Both WT-Greedy-R and CT-Greedy-R need to globally select protectors with multiple times for every target and also has high time consumption.

\emph{Utility analysis}. Privacy preserving often reduces the graph utility which is very critical for a released graph. Currently, the graph utility is generally represented by many important statistical graph metrics (see Table \ref{tab1}).

\begin{table}[htb]
\caption{The main metrics for the graph utility analysis.}
\begin{center}
\begin{tabular}{|c|c|}
\hline
 Metric Notation & Description \\
\hline
 $\overline{l}$ & The average path length for all node pairs \\
 \hline
 $\overline{clust}$ & The average clustering coefficient \cite{ref18} \\
 \hline
 $r$ & The assortativity coefficient \cite{ref37} \\
 \hline
 $\overline{cn}$ & The average core number of all nodes \cite{ref39} \\
 \hline
 $\mu$ & The second largest eigenvalue \cite{ref6} \\
 \hline
 $Mod$ & The modularity metric of community structures \cite{ref38}\\
\hline
\end{tabular}
\label{tab1}
\end{center}
\end{table}

1) \emph{Distance}. Distance between any pair of nodes $(u, v)$ is a critical metric to evaluate end-to-end characteristic of the graph, denoted by $l_{uv}$ which is the number of passing through links from node $u$ to $v$. The average path length of a graph is $\overline{l}= \sum_{u,v \in V,u \neq v}l_{uv}/(N(N-1)/2)$.

2) \emph{Clustering Coefficient}. Clustering coefficient \cite{ref18} of a node in a graph quantifies how close its neighbors are to being a clique, which is mathematically defined by the notation $clust_v= |\{w:w \in \Gamma_u \cap \Gamma_v\}|/(d_v (d_v-1)/2)$, where $\Gamma(v)$ is the set of neighboring nodes of $v$. Then the average clustering coefficient can be computed by $\overline{clust} = \sum_{v\in V} clust_v/N$.

3) \emph{Assortativity coefficient}. Assortativity \cite{ref37} (or assortative mixing) is a preference for a network's nodes to attach to others that are similar in some way, quantifying the degree correlations. Assuming $\varphi_{ij}$ as the probability to find a node with degree $i$ and degree $j$ at the two ends of a randomly selected edge, and $q_i$ as the probability to have a degree $i$ node at the end of a link, the assortativity coefficient is $r=\sum_{ij}ij(\varphi_{ij}-q_i q_j)/\sigma_{q^2}$, where $\sigma_{q^2}$ is the variance of the distribution $q_i$.

4) \emph{Core number}. The core number of a node $v$ indicates the node importance by k-shell decomposition \cite{ref39}, denoted by $cn_v$, and average core number of all nodes is $\overline{cn} = \sum_{v\in V} cn_v/N$.

5) \emph{Eigenvalue centrality}. The Laplacian matrix is used to find many useful properties of a graph, $L=D-G$, where the matrix $D$ is the degree matrix in which $D_{vv}=d_v$, and otherwise $D_{uv} = 0 (u \neq v)$. The eigenvector of Laplacian matrix is $\mu=\{\mu_1, \mu_2, \cdots ,\mu_N\}$, and let $\mu$ represents the second largest eigenvalue which can well describe the graph property \cite{ref2}.

6) \emph{Modularity}. Community modularity \cite{ref38} of a network is critical for describe the structure organizations for real relationship establishments. The community property of a graph is evaluated by the modularity metric which is denoted by $Mod=1/2m \sum_{i,j} [G_{ij}-d_i d_j/2m] \delta(c_i,c_j)$, where $G_{ij}=1$ if node $i$ and $j$ are connected, $d_i$ is the degree of node $i$, $c_i$ is the community number to which vertex $i$ is assigned, and $m$ is the total number of links in the current graph. $\delta$-function $\delta(u,v)$ is 1 if $u=v$ and 0 otherwise.

To analyze the utility loss between the original graph and a released graph, we use the utility loss ratio defined by
\begin{displaymath}
ulr(z, G, G')=|\frac{(z(G)-z(G'))}{z(G)}|
\end{displaymath}
where $z(G)$ and $z(G')$ represents a metric (listed in Table \ref{tab1}) for original and perturbed graphs respectively.  The average utility loss ratio for all utility metrics is
\begin{displaymath}
\overline {ulr}(G, G') = \frac{\sum_z ulr(z,G,G')}{\# ~of ~utility ~metrics}
\end{displaymath}

In general, we hope all targets are perfectly protected such that the adversaries can't infer the existence of targets. This can be achieved by deleting a protector set $P$ which can result in the total similarity to 0, namely $\sum_{t \in T}s(P,t) = 0$. We treat it as full protection for all targets.
\begin{table}[htb]
\newcommand{\tabincell}[2]{\begin{tabular}{@{}#1@{}}#2\end{tabular}}
\centering
\caption{Comparisons of utility loss ratio for all greedy algorithms on \emph{Arenas-email}, with $|T|=20$.}
\label{tab2}
\begin{tabular}{|l|l|l|l|l|l|l|}
\hline
\multirow{2}{*} & \multirow{2}{*}{$G\setminus T$} & \multirow{2}{*}{\tiny{\tabincell{c}{SGD-\\Greedy(-R)}}} & \multicolumn{2}{|c|}{CT-Greedy(-R)} & \multicolumn{2}{|c|}{WT-Greedy(-R)} \\
\cline{4-7}
 &  &  & DBD & TBD & DBD & TBD \\
\hline
\emph{Triangle} & 0.64\% & 1.95 \% & 1.95 \% & 1.95 \% & 1.95 \% & 1.95 \% \\
\hline
\emph{Rectangle} & 0.64 \% & 2.49 \% & 2.53 \% & 2.47 \% & 2.60 \% & 2.38 \% \\
\hline
\emph{RecTri} & 0.64 \% & 1.23 \% & 1.28 \% & 1.27 \% & 1.28 \% & 1.28 \% \\
\hline
\end{tabular}
\end{table}
\begin{table}[htb]
\newcommand{\tabincell}[2]{\begin{tabular}{@{}#1@{}}#2\end{tabular}}
\centering
\caption{Comparisons of utility loss ratio for all greedy algorithms on \emph{Arenas-email}, with $|T|=50$.}
\label{tab3}
\begin{tabular}{|l|l|l|l|l|l|l|}
\hline
\multirow{2}{*} & \multirow{2}{*}{$G\setminus T$} & \multirow{2}{*}{\tiny{\tabincell{c}{SGD-\\Greedy(-R)}}}  & \multicolumn{2}{|c|}{CT-Greedy(-R)} & \multicolumn{2}{|c|}{WT-Greedy(-R)} \\
\cline{4-7}
 &  &  & DBD & TBD & DBD & TBD \\
\hline
\emph{Triangle} & 1.14 \% & 2.97 \% & 2.97 \% & 2.97 \% & 2.97 \% & 2.97 \% \\
\hline
\emph{Rectangle} & 1.14 \% & 7.98 \% & 8.63 \% & 7.93 \% & 7.97 \% & 8.64 \% \\
\hline
\emph{RecTri} & 1.14 \% & 3.14 \% & 3.54 \% & 3.15 \% & 3.46 \% & 3.11 \% \\
\hline
\end{tabular}
\end{table}

In Table \ref{tab2} and Table \ref{tab3}, we set $|T|=20$ and $|T|=50$ respectively for graph \emph{Arenas-email}. For any specific subgraph pattern, all greedy algorithms in this work can achieve the goal of the full protections with very little utility loss. The protection for \emph{Rectangle} pattern needs deleting a higher number of protectors and causes a bit higher utility loss under every greedy protection mechanism. With a higher number of targets, more protectors are need for full protections at the cost of a bit higher utility loss. Taking \emph{Triangle} pattern for instance, fully protecting 20 targets lead to 1.95\% utility loss, while protecting 50 targets causes 2.97\% utility loss under the SGB-Greedy(-R) algorithm.
\begin{table}[htb]
\newcommand{\tabincell}[2]{\begin{tabular}{@{}#1@{}}#2\end{tabular}}
\centering
\caption{Comparisons of utility loss ratio for all greedy algorithms on \emph{DBLP}, with $|T|=52$.}
\label{tab4}
\begin{tabular}{|l|l|l|l|l|l|l|}
\hline
\multirow{2}{*} & \multirow{2}{*}{$G\setminus T$} & \multirow{2}{*}{\tiny{\tabincell{c}{SGD-\\Greedy(-R)}}}  & \multicolumn{2}{|c|}{CT-Greedy(-R)} & \multicolumn{2}{|c|}{WT-Greedy(-R)} \\
\cline{4-7}
 &  &  & DBD & TBD & DBD & TBD \\
\hline
\tiny{\emph{Triangle}} & 0.011\% & 0.014\% & 0.017\% & 0.017\% & 0.018\% & 0.018\% \\
\hline
\tiny{\emph{Rectangle}} & 0.011\% & 0.016\% & 0.015\% & 0.015\% & 0.018\% & 0.018\% \\
\hline
\tiny{\emph{RecTri}} & 0.011\% & 0.012\% & 0.016\% & 0.016\% & 0.013\% & 0.013\% \\
\hline
\end{tabular}
\end{table}

In Table \ref{tab4}, we show the utility loss for all methods on \emph{DBLP} graph. Because the \emph{DBLP} graph is huge, and many utility metrics such as the average path length and eigenvalue can't be efficiently computed on a general sever. Here we show the results of clustering coefficient and core number, and the results show that the graph utility loss is very tiny if a small number of targets are protected by deleting a limited budget of link deletion $k=25$.

The experimental results show that protecting key targets in large social graphs is workable. Due to personal considerations, if two users want to hide the link between them, in order that the adversaries can't infer the link existence by analyzing a specific subgraph pattern that the two users frequently participate in, our work can do efficient assistance.
\subsection{Extended Discussion}
To extend, a fully protected graph can defend a serial of target subgraph related link predictions, for instance, the \emph{Triangle} based predictions including Jaccard \cite{ref19}, Salton\cite{ref20}, S{\o}rensen\cite{ref21}, Hub Promoted \cite{ref22}, Hub Depressed \cite{ref23}, Leicht-Holme-Newman \cite{ref23}, Adamic-Adar \cite{ref24}, and Resource Allocation \cite{ref25}, in which the prediction probability for every target is 0.

However, if the dissimilarity employs the above-mentioned metrics, the theoretical proofs for monotonicity and submodularity are not satisfied. That is why we choose subgraph pattern based dissimilarities. In the following, we discuss why these mentioned metrics don't have theoretical guarantees and why the link addition and link switch are not workable.

\emph{Illustrations for other triangle related dissimilarity metrics}. We investigate the non-scalability in the link deletion process of TPP for many other triangle related similarity indices \cite{ref19,ref20,ref21,ref22,ref23,ref24,ref25} including the Jaccard, Salton, S{\o}rensen, Hub Promoted, Hub Depressed, Leicht-Holme-Newman, Adamic-Adar, and Resource Allocation.
\begin{figure}[ht!] 
\centering
\includegraphics[width=3.2in]{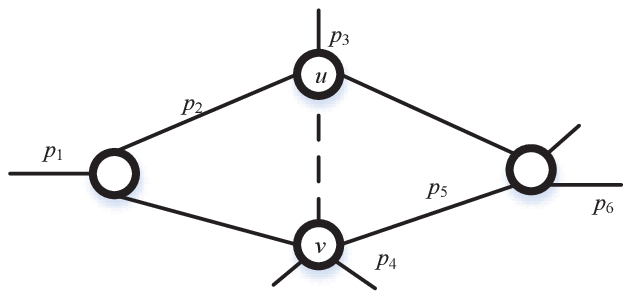}
\caption{An example for illustrations of the proof of other dissimilarity indices including the Jaccard, Salton, S{\o}rensen, Hub Promoted, Hub Depressed, Leicht-Holme-Newman and Adamic-Adar. For simplicity, we assume there is only one target $T =\{(u,v)\}$. }
\label{fig7}
\end{figure}

1) Jaccard \cite{ref19}. The similarity for node pair $(u,v)$ is defined by $|\Gamma_u \cap \Gamma_v |/|\Gamma_u \cup \Gamma_v |$, where $\Gamma_u$ is the neighbor set of node $u$, and the dissimilarity function is $f(P,T)=1-|\Gamma_u \cap \Gamma_v |/|\Gamma_u \cup \Gamma_v |$. As shown in Fig.~\ref{fig7}, initially, $f(\emptyset,T)=1-2/5$. Firstly, we discuss the monotonicity of this dissimilarity function. Cases: a) only delete protector $p_1$, namely $P=\{p_1\}$, $f(P,T)=1-2/5=f(\emptyset,T)$; b) $P=\{p_2\}$, $f(P,T)=1-1/5 > f(\emptyset,T)$; c) $P=\{p_3\}$, $f(P,T)=1-2/4 < f(\emptyset,T)$. It can be inferred that the monotonicity can't be guaranteed, so the greedy algorithm can not achieve near optimal results.

2) Salton \cite{ref20}. It is defined by $|\Gamma_u \cap \Gamma_v |/\sqrt{d_u d_v}$, where $d_u$ is the degree of node $u$, and the dissimilarity function is $f(P,T)=1-(|\Gamma_u \cap \Gamma_v |)/\sqrt{d_u d_v}$. For the target link, the initial dissimilarity is $f(\emptyset,T)=1-2/\sqrt{12}$. The cases for deleting one protector: a) $P=\{p_1\}$, $f(P,T)=1-2/\sqrt{12} = f(\emptyset,T)$; b) $P=\{p_2\}$, $f(P,T)=1-1/\sqrt{8} > f(\emptyset,T)$; c) $P=\{p_3\}$, $f(P,T)=1-2/\sqrt{8} < f(\emptyset,T)$. The monotonicity can't be satisfied.

3) S{\o}rensen \cite{ref21}. The similarity is defined by $2|\Gamma_u \cap \Gamma_v |/(d_u + d_v)$, and the dissimilarity function is $f(P,T)=1- 2|\Gamma_u \cup \Gamma_v|/(d_u+d_v)$, and $f(\emptyset,T)=1-4/7$. Cases: a) $P = \{p_1\}$, $f(P,T)=1-4/7  = f(\emptyset, T)$; b) $P=\{p_2\}$, $f(P,T)=1-2/6 > f(\empty,T)$; c) $P=\{p_3\}$, $f(P,T)=1-4/6 < f(\emptyset,T)$. The monotonicity can't be satisfied.

4) Hub Promoted (HP) \cite{ref22}. Defined by $|\Gamma_u \cap \Gamma_v |/\min\{d_u, d_v\}$, the dissimilarity function is $f(P,T)=1-|\Gamma_u \cap \Gamma_v |/\min\{d_u,d_v\}$, and $f(\emptyset,T)=1-2/3$. Cases: a) $P=\{p_1\}$, $f(P,T)=1-2/3=f(\emptyset,T)$; b) $P=\{p_2\}$, $f(P,T)=1-1/2 > f(\emptyset,T)$; c) $P=\{p_3\}$, $f(P,T)=1-2/2<f(\emptyset,T)$. The monotonicity can't be satisfied.

5) Hub Depressed (HD) \cite{ref23}. Defined by $(|\Gamma_u\cap \Gamma_v |)/(\max \{d_u,d_v\})$, the dissimilarity function is $f(P,T)=1-(|\Gamma_u \cap \Gamma_v |)/(\max \{d_u, d_v\})$, and $f(P,\emptyset)=1-2/4$. Cases: a) $P=\{p_1\}$, $f(P,T)=1-2/4=f(\emptyset,T)$; b) $P=\{p_2\}$, $f(P,T)=1-1/3>f(\emptyset,T)$; c) $P=\{p_4\}$, $f(P,T)=1-2/3 < f(\emptyset,T)$. The monotonicity can't be satisfied.

6) Leicht-Holme-Newman (LHN) \cite{ref23}. The similarity is defined by $|\Gamma_u \cap \Gamma_v |/d_u d_v$, and the dissimilarity function is $f(P,T)=1-|\Gamma_u \cap \Gamma_v |/d_u d_v$, and $f(\emptyset,T)=1-2/12$. One protector selecting cases are: a) $P=\{p_1\}$, $f(P,T)=1-2/12=f(\emptyset,T)$; b) $P=\{p_2\}$, $f(P,T)=1-1/8> f(\emptyset,T)$; c) $P=\{p_3\}$, $f(P,T)=1-2/8 < f(\emptyset,T)$. The monotonicity can't be satisfied.

7) Adamic-Adar (AA) \cite{ref24}. The AA similarity is defined by $\sum_{v'\in \Gamma_u \cap \Gamma_v}\frac{1}{\log{d_{v'}}}$, and the dissimilarity function is $f(P,T)=C-\sum_{v'\in \Gamma_u \cap \Gamma_v}\frac{1}{\log{d_{v'}}}$, and $f(\emptyset,T)=C-(\frac{1}{\log{3}} + \frac{1}{\log{4}})$. Cases: a) $P=\{p_1\}$, $f(P,T)=C-(\frac{1}{\log{2}} +\frac{1}{\log{4}})<f(\emptyset,T)$; b) $P=\{p_2\}$, $f(P,T)=C-\frac{1}{\log{4}} >f(\emptyset,T)$. The monotonicity can't be satisfied.
\begin{figure}[ht!] 
\includegraphics[width=3.2in]{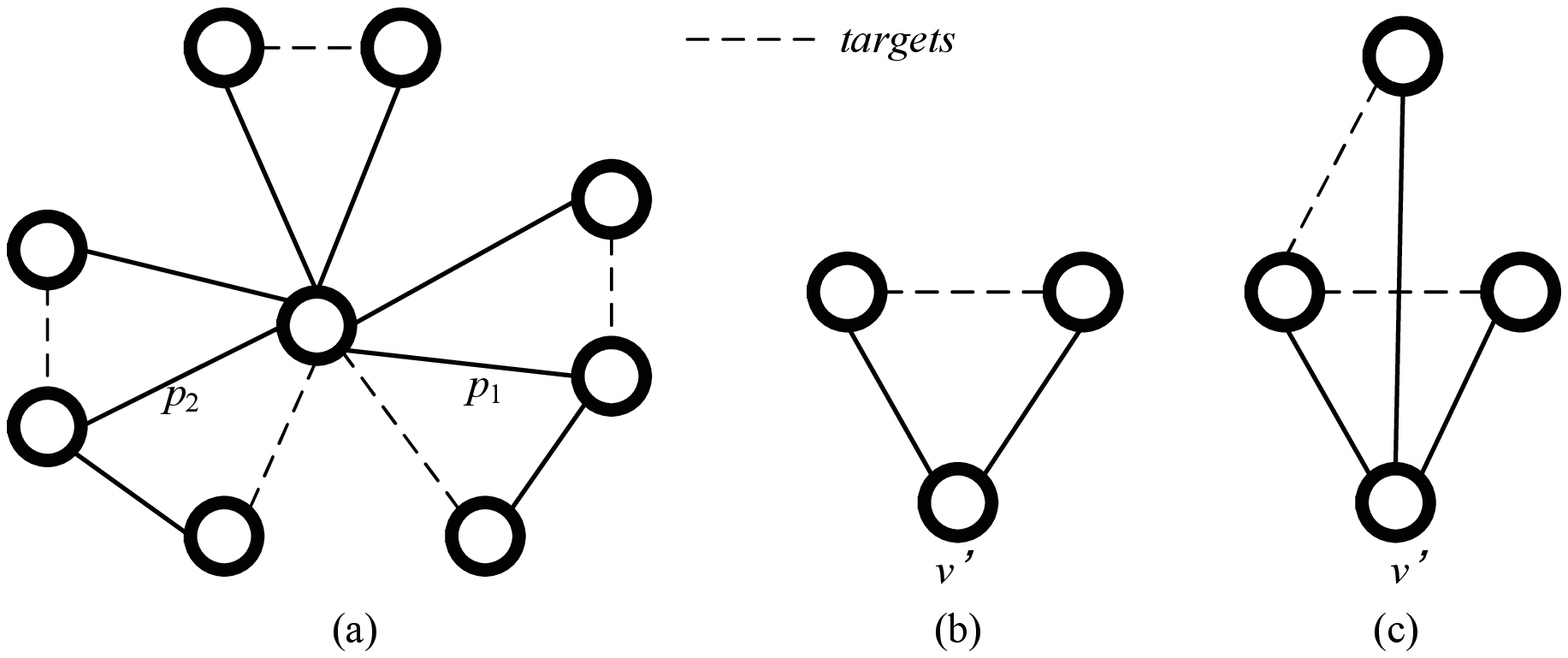}
\caption{An example for the proof illustration of Resource Allocation based dissimilarity index.  }
\label{fig8}
\end{figure}

8) Resource Allocation (RA) \cite{ref25}. The similarity is defined by $\sum_{v'\in \Gamma_u \cap \Gamma_v}\frac{1}{d_{v'}}$, and dissimilarity function is $f(P,T)=C-\sum_{v'\in \Gamma_u \cap \Gamma_v}\frac{1}{d_{v'}}$, and $f(\emptyset,T)=C-(1/3+1/4)$. Cases for random protector deletion include: a) $P=\{p_1\}$, $f(P,T)=C-(1/2+1/4)<f(\emptyset,T)$; b) $P=\{p_2\}$, $f(P,T)=C-1/4>f(\emptyset,T)$. The monotonicity can't be satisfied.

Furtherly, we assume all protectors are iteratively sorted by the number of participations in target triangles, and at every step the protector of the highest value is selected. However, the monotonicity or the submodularity can not be guaranteed under the Jaccard, Salton, S{\o}rensen, Hub Promoted, Hub Depressed, Leicht-Holme-Newman, Adamic-Adar, and Resource Allocation similarity indices. Here we take the RA similarity as an example which is a bit more complex than others.

For multiple targets, the dissimilarity function of RA is defined by $f(P,T)=C-\sum_{(u,v)\in T}\sum_{v'\in \Gamma_u \cap \Gamma_v}\frac{1}{d_{v'}} $.

\emph{Lemma} 6. The Resource Allocation based dissimilarity function $f(P,T)$ is monotone. For any sets $A\subseteq B\subseteq P$, assume the protectors in $P$ is iteratively selected as the link of the highest participations in all target triangles, then we have $f(A,T)\le f(B,T)$.

\emph{Proof}. Deleting a protector $p$, only the score of the targets (as in Fig.~\ref{fig8}(b)) whose target subgraphs include the protector will change. We assume $\theta$ is the number of targets whose two end nodes have common neighbor $v'$. For instance, $\theta=1$ and $\theta=2$  in Fig.~\ref{fig8}(b) and Fig.~\ref{fig8}(c) respectively. For any node $v'$, initially, the similarity score is $\theta / d_{v'} >0$. Deleting a protector adjacent to $v'$, the degree of $v'$ reduces to $d_{v'}-1$, and one or more (\emph{i.e.} $y$) target triangles are broken. Then for $v'$, the score is $(\theta-y)/(d_{v'}-1)$. Due to $y\ge 1$, then $(\theta - y)/(d_{v'}-1)<\theta/d_{v'}$  (\emph{e.g.} in Fig.~\ref{fig8}(b), deleting any link will cause the similarity score to be 0, less than the value 1/2 before deletion), the similarity for $v'$ decreases, while the total dissimilarity will increase. The monotonicity is satisfied.

We tried to prove the submodularity for the RA based dissimilarity function. For an example, in Fig.~\ref{fig8}(a), we assume the sets $A=\emptyset, B=\{p_1\}$, $A\subseteq B$ and $p=p_2$. By the dissimilarity function, initially, we have $f(\emptyset,T)=C-(3/6+1)$; $f(B,T)=C-(2/5+1/2)$; $f(B\cup \{p\},T)=C-1/4$, and $f(A\cup \{p\},T)=C-(2/5+1/2)$. In this case, we have $\Delta f(B,T)=f(B\cup \{p\},T)-f(B,T)=2/5+1/4$, and $\Delta f(A,T)=f(A\cup \{p\},T)-f(A,T)=3/5$. It can be seen $\Delta f(A,T) < \Delta f(B,T)$. The submodularity is not satisfied.

\emph{Illustrations for Link Additions}. We define the dissimilarity function by $f'(P',T)=C-\sum_{t\in T}s'(P',t)$ where $P'$ is the set of added links, and $s'(P',t)$ is the similarity score or the number of target subgraphs for target $t$. In fact, adding a new protector $p$ into the graph will never break the existing target subgraphs. Then we have $s'(P'\cup \{p\},t) \ge s'(P',t)$ for any target $t$ in any subgraph pattern, namely $f'(P',T)\le f'(P'\cup \{p\},T)$ which indicates that the dissimilarity function is not an increase function. The monotonicity property is not satisfied. Then it is not necessary to check the submodularity property of the objective dissimilarity function.

\emph{Illustrations for Link Switching}. We define the dissimilarity function as $f^{\#}(P^{\#},T)=C-\sum_{t\in T}s^{\#}(P^{\#},t)$ where set $P^{\#}$ is the switched links, and $s^{\#}(P^{\#},t)$ is the similarity score or the number of target subgraphs for target $t$ when switching a set $P^{\#}$  of links in the graph. Here we assume the switching process is totally random. In general, a link switching procedure can be accomplished by two steps: 1) Randomly delete $k$ existing links from the original graph; 2) Randomly add $k$ new links between the unconnected nodes' pairs of the graph. If we first randomly select a link beyond any target subgraphs, and then add a new link between a pair of unconnected nodes. As discussed above, the dissimilarity score might decrease. The monotonicity is not satisfied.

Moreover, we need to do more effort on the applications (\emph{e.g} fraud detection \cite{sun2020kollector,sun2017contaminant,li2018significant}) of the proposed privacy preserving mechanism.


%


\section{Conclusion and Future Work}
To summarize, in this paper, we studied a novel target privacy preserving problem which accurately focused on the protection of important, private and easily attacked targets. Only deleting all targets from the graph was insufficient to defend the adversarial link predictions, and a set of protectors were intensively eliminated to promote the attack defense ability of all targets. With limited deletion budget, the optimal protector selection is the key issue. Firstly, an objective dissimilarity function was defined, where higher dissimilarity scores means better protection. Secondly, we proved the optimal protector selection is NP-hard, and then designed near optimal solutions for two budget assignment scenarios: 1) single global budget, where the SGB-Greedy algorithm can achieve at least $1-1/e$ approximation; 2) multiple local budgets, where the CT-Greedy and WT-Greedy algorithm achieves an approximation ratio $1/2$ and 0.46 respectively. Thirdly, scalable implementations were done to improve the running efficiency of proposed greedy algorithms. Finally, the experimental results showed that all proposed greedy algorithms can fully protect all targets at the cost of a bit utility loss. Our work is general and can be used for any subgraph pattern based privacy preserving.

We further explored many other similarity metrics as a part of dissimilarity functions and tried to prove their monotonicity and submodularity. We found that the dissimilarity function which was defined by many widely used similarity metrics such as Jaccard \cite{ref19}, Salton\cite{ref20}, S{\o}rensen\cite{ref21}, Hub Promoted \cite{ref22}, Hub Depressed \cite{ref23}, Leicht-Holme-Newman \cite{ref23}, Adamic-Adar \cite{ref24}, and Resource Allocation \cite{ref25} is not monotone or submodular. In other words, the proposed greedy algorithms of this work can't be directly used to achieve near optimal results for them, but the fully protected graphs of our methods can defend all above adversarial link predictions. Furthermore, we analyzed that the dissimilarity function of this work doesn't satisfy the monotonicity or submodularity for link addition and link switch mechanisms.

To our best knowledge, this work is the first to focus on target privacy preserving problem which is still open and challenging in real social graphs: 1) more TPP mechanisms against kinds of other link predictions (\emph{e.g.} Katz \cite{katz} index based prediction); 2) target node privacy preserving technologies; 3) applications into real trust systems or social graphs.

\section*{Acknowledgments}
This work was supported by the Key R\&D Program of Shaanxi Province of China (No. 2019ZDLGY12-06), NSFC (No. 61502375, 61672399, U1405255), the China 111 Project (No. B16037), Shaanxi Science \& Technology Coordination \& Innovation Project (No. 2016TZC-G-6-3), and NSF (No. III-1526499, III-1763325, III-1909323, CNS-1930941, CNS-1626432).




\bibliographystyle{IEEEtran}
\bibliography{ref}

\end{document}